\newcommand*\pFqskip{8mu}
\catcode`,\active
\newcommand*\pFq{\begingroup
        \catcode`\,\active
        \def ,{\mskip\pFqskip\relax}%
        \dopFq
}
\catcode`\,12
\def\dopFq#1#2#3#4#5{%
        {}_{#1}F_{#2}\biggl[\genfrac..{0pt}{}{#3}{#4};#5\biggr]%
        \endgroup
}
\documentclass[showpacs,showkeys,preprint,aps]{revtex4}
\usepackage{todonotes}
\usepackage{graphicx}
\usepackage{dcolumn}
\usepackage{bm}
\usepackage{comment}
\usepackage{hyperref}

\usepackage{amsfonts,amsmath,amssymb,cancel}
\usepackage{todonotes}

\begin{document}


\title{Molecular analogue for scalar dynamics in an AII metric}

\author{Davi M. E. Moreira}
\author{Matheus E. Pereira}
\author{Alexandre G. M. Schmidt}%
 \email{agmschmidt@id.uff.br}
\affiliation{%
  Instituto de Ci\^encias Exatas, Universidade Federal Fluminense,
  \\ 27213-145 Volta Redonda --- RJ, Brazil
}%
\affiliation{Programa de P\'os Gradua\c{c}\~ao em F\'isica, Instituto de F\'\i sica, Universidade Federal Fluminense,\\
	24210-346 Niter\'oi --- RJ, Brazil}




\date{\today}
\begin{abstract}

Tachyons are hypothetical particles that propagate faster than light, yet they have never been observed in nature or in the laboratory. In this work, we introduce the $H_2^+$ molecule as an analogue for the dynamics of a spinless test particle interacting with the gravitational field generated by a tachyon. The tachyonic spacetime is modeled using an AII metric, and the problem is analyzed through the Klein–Gordon equation for a scalar field in this background. We compute the quasi-normal modes and the Hawking radiation spectrum associated with the system. By introducing a specific external potential --- defined precisely for this purpose and fine-tuned through a careful comparison of system constants --- we demonstrate that both the radial and angular components of a test particle's wave function can be effectively reproduced by the electronic dynamics within $H_2^+$. We thus propose a mathematical framework for studying a molecular analog model of an extreme gravitational system.

\end{abstract}

\keywords{General Relativity; Analogue models; Tachyons; Klein-Gordon equation; Heun functions}
\maketitle
\section{Introduction}
\label{Intro}
    Ehlers and Kundt \cite{ehlers}, in their investigation of static gravitational fields in the vacuum, introduced and classified the Petrov-type D fields, among which are what they called A and B metrics. These metrics are divided into three subclasses, namely, I, II, and III. The metric labeled as AI is the well-known Schwarzschild solution for a spherically symmetric mass distribution, meanwhile the other two lacked a physical interpretation and were studied as topological black holes \cite{griffithsepodolsky}. 

    Peres \cite{peres}, in 1970, realized that if a superluminal speed boost is performed in the source of the Schwarzschild metric, written in isotropic coordinates, one can obtain the AII metric (and also the BI, that complements the spacetime). The former fact reveals that a suitable physical interpretation is that it describes the gravitational field of a tachyon: a particle that travels faster than light through spacetime with a -- possibly infinite -- velocity $v>c$. Such an object would have a spacelike worldline, and the metric in question describes only the region of the spacetime that is causally related to the tachyon \cite{gott}. 
    
    Hruska and Podolsky \cite{podolsky} explained clearly the process of boosting the Schwarzschild metric to infinite speed to obtain both AII and BI metrics that, together, can describe the entire spacetime. By their approach, the region of validity of the metric and the Mach-Cherenkov gravitational shockwave arises naturally, giving us the correct boundary conditions to the problem of particle motion in this context. As well as them, Gott \cite{gott} has described in detail that timelike geodesics, i.e. the trajectory of particles that have mass, can not surpass the region limited by the shockwave, and this fact is due to an infinite discontinuity in the Weyl curvature scalar \cite{podolsky}.
    
    Despite the interesting phenomena surrounding the system, the great wealth of details, and the theoretical interest, its deficiency lies in the physical infeasibility. Without taking into account the discussion of whether tachyons can even exist, these superluminal particle were never observed in nature nor realized in laboratory. Despite this, there is a mechanism that allows us to investigate this inaccessible physical structure, as well as many other problems --- such as magnetic monopoles, dyons, wormholes, black strings and so forth --- , in the laboratory: analogue models. 
    

    The search for replicating the behavior of gravitational systems is of great interest in contemporary physics, see \cite{carla,schmidt,schmidt2}, with the pursuit of the enhancement of this research programme. However, this story has begun long ago with Zel'dovich's first analogue model \cite{zeldovich}. His experimental work was able to emulate the dynamics of a black hole to elucidate the functioning of the Penrose process, and even predict a new effect that would appear in this context. Furthermore, other analogies came up, just as Unruh's \cite{unruh} sonic black hole that could replicate the Schwarzschild black hole using sound waves in a fluid, and many others. Unruh's analogy also manifested a Hawking-like effect, the core motivation of the analogue gravity research in the 2000s \cite{fischer-visser, richartz-cqg, rotating-bath-tub, trapped-ions}. 
    
    At the present, this research programme became wider, embracing the most diverse areas of physics, such as hydrodynamical \cite{fifer2019analog, torres2020quasinormal}, optical \cite{tinguely2020optical}, condensed matter systems \cite{mertens2022thermalization, de2022generalized, furtado-catenoide, surface-plasmon-Smolyaninov}, Bose-Einstein condensates \cite{lee-tachyonic-BEC, quasi-1D} and gauge theories \cite{grassi, geometrodynamics}. This new phase of the programme has novel purposes that go beyond the original goal of mimicking the Hawking effect. It sheds light on new gravitational phenomena \cite{nonlinear-effects} and can even predict as well as verify experimentally new quantum effects \cite{carla, nature-super-radiance}, and is at this point where the present work lies at. 
    

The objective of our work is to present an analogue model for the dynamics of a scalar particle propagating in the gravitational field of a tachyon. In our model, the target system is a scalar test-particle interacting with a gravitational field produced by a tachyon. The laboratory system is a spinless charged particle that interacts with two static charges plus a certain very specific external potential. The most simple example of such laboratory system is the hydrogen molecular ion in the presence of this external potential. It is worth noting that the main distinction, among others, between this newly proposed molecular analogue model and the most popular ones, the hydrodynamic ones, is in the degrees of freedom: while the latter is restricted to two degrees of freedom, the former has no restrictions in this regard. However, the price to be paid for this type of analogy construction is the fact that such an external potential may not be exactly achievable in the laboratory, potentially requiring approximations and potential engineering to realize it. The singularities of a differential equation are the keypoint of our analogue model. More specifically, the quantum dynamics of the test particle in the target system is governed by the Klein--Gordon equation, which after separation results in the confluent Heun's equation \cite{ronveaux,zamp2}. On the other hand, the test-particle in the laboratory system obeys the Schr\"odinger equation \cite{wilson}, which after variable separation also yields a confluent Heun's equation. The radial differential equations, \eqref{radialescala} and \eqref{h2+potencial}, of both systems are the same. In addition, the angular differential equations, \eqref{angularmelhor} and \eqref{angularanalogo}, are also the same. So, in our analogue model we can probe the semi-classical dynamics of a quantum particle interacting with such an extreme gravitational field by precisely adjusting the physical quantities of the laboratory system according to the mappings given by equations \eqref{analogia-parametros-eq-radial} and \eqref{analogia-parametros-eq-angular}.

    The outline for our paper is the following: in section \ref{sec:Klein-Gordon} the Klein-Gordon equation is presented for the AII metric and the separation of variables is achieved. Section \ref{sec:Solucoes} is dedicated to the solution of the angular and radial parts of the problem, respectively, where we obtain the appropriated values for the separation constant and energy eigenvalues. In section \ref{sec:Hawking}, the Damour-Ruffini-Sannan method is applied in order to obtain the radiation spectrum and Hawking temperature associated with the event horizon of the metric, and quasi-normal modes related to the tachyonic singularity are calculated and presented. The analogue models of radial and angular dynamics are studied in sections \ref{sec:Analogo Radial} and \ref{sec:Analogo Angular}, respectively. Finally, we conclude the work with final remarks in section \ref{sec:Conclusao}.  

    \section{Klein-Gordon equation}\label{sec:Klein-Gordon}

    Let us consider the AII metric \cite{podolsky}
    \begin{equation}
        ds^2=-\left(\frac{b}{z}-1\right)dt^2+\left(\frac{b}{z}-1\right)^{-1}dz^2+z^2(dr^2+\sinh^2{r}d\varphi^2)
        \label{metrica}
    \end{equation}
    where $b$ is a continuous parameter that will be interpreted as the absolute value of the mass of the tachyon. In these coordinates, $z$ is the radial-like degree of freedom, restricted to the range $\left[0,\infty\right)$, $r \in \left[0,\infty\right)$ and $\phi \in \left[0,2\pi\right)$. The quantum dynamics of a scalar particle with mass $\mu_0$ in the background of this tachyonic singularity is governed by the Klein-Gordon equation,
    \begin{equation}
        \frac{1}{\sqrt{-g}}\partial_\mu\left(g^{\mu\nu}\sqrt{-g}\partial_\nu \Psi\right)-\mu_0^2\Psi=0,
        \label{kleingordon}
    \end{equation}
    where $g=\det(\Tilde{g})$ denotes the determinant of the metric tensor, which components can be straightforwardly read from the line element \eqref{metrica} and with units so that $G=c=\hbar=1$. By introducing the \textit{ansatz},
    \begin{equation}
        \Psi(t,z,r,\varphi)=e^{im\varphi}e^{-i\sigma t}R(r)Z(z),
    \end{equation}
    and a separation constant $\lambda$, equation \eqref{kleingordon} separates into what is the radial-like equation
    \begin{equation}
        \frac{d}{dz}\left[ z(b-z)\frac{dZ}{dz} \right] + \left[ z^2\left(\frac{b}{z}-1\right)^{-1}\sigma^2-\mu_0^2z^2-\lambda \right]Z=0,
        \label{radial}
    \end{equation}
and the angular-like equation
    \begin{equation}
        \frac{1}{\sinh{r}}\frac{d}{dr}\left( \sinh{r}\frac{dR}{dr} \right) + \left( \lambda -\frac{m^2}{\sinh^2{r}} \right) R=0,
        \label{angular}
    \end{equation}
in this coordinate system, where $m$ is an integer and $\sigma$ is a complex number, directly related to the energy eigenvalues. Thus, these are the radial-like and angular-like differential equations for our target system, the scalar particle in a tachyonic gravitational field. They can model physical systems that appear very different from the one illustrated in this section. 


\section{Exact solutions}\label{sec:Solucoes}

    The angular-like equation \eqref{angular} can be immediately transformed into the associated Legendre equation by introducing the new variable $x=\cosh{r}$. Proceeding this way, the solutions are
    \begin{equation}
        R(r)=c_1 P^m_\xi(\cosh{r}) + c_2Q^m_\xi(\cosh{r}),
    \end{equation}
    where
    \begin{equation}
        \xi=\frac{-1\pm\sqrt{1-4\lambda}}{2}.
        \label{equacaoxi}
    \end{equation}
    However, since $\xi$ is not necessarily an integer, and $r\geq 0$, one must use the correct representation of the Legendre associated function $P_\xi^m(w)$, with $w=\cosh r$, which is given in terms of Gauss's hypergeometric function $_2F_1$ \cite{bateman,magnus},
    \begin{equation}
        P_\xi^m(w)=\frac{1}{\Gamma(1-m)}\left(\frac{w+1}{w-1}\right)^{m/2}\left(\frac{w+1}{2}\right)^{\xi} \pFq{2}{1}{-\xi,,-\xi-m}{1-m}{\frac{w-1}{w+1}},
        \label{2f1continuacaoanalitica}
    \end{equation}
where $\Gamma(z)$ is the well-known gamma function.

    As stated by Gott \cite{gott} and explained by Hruska and Podolsky \cite{podolsky}, the particle should experience a gravitational shock wave -- which is interpreted as the Mach-Cherenkov cone produced by the superluminal source -- that prohibits its motion through it. In our coordinate system, this cone is represented by $r\to\infty$. Therefore, in order to find the allowed values of the separation constant $\lambda$, we impose that the ``angular'' part of the wavefunction vanishes in this limit
    \begin{equation}
        \lim_{w\to\infty}P_\xi^m(w)=0.
    \end{equation}
    By taking the limit using the property of Gauss' hypergeometric function \cite{bateman}
    \begin{equation}
        \pFq{2}{1}{a,b}{c}{1} = \frac{\Gamma(c)\Gamma(c-a-b)}{\Gamma(c-a)\Gamma(c-b)}
    \end{equation}
    we conclude that there is a restriction of the parameters $\xi$ and $m$ to the negative integers and positive integers, respectively, in order to match the required boundary condition. By inverting equation \eqref{equacaoxi}, employing its negative root --- otherwise $\lambda$ would be complex and the vanishing of the wavefunction at the ``gravitational barrier" could not be guaranteed --- and setting $\xi=-n,\quad n \in \mathbb{N}^*$, the possible values of $\lambda$ are $\lambda=-n(n-1)$, i.e., $\lambda= 0,-2,-6,-12,-20,-30,-42,...$ and so forth.
    

Turning our attention to the radial-like part of the wavefunction, we rewrite the equation \eqref{radial} as 
    \begin{equation}
        Z''+\left( \frac{1}{z}+\frac{1}{z-b} \right)Z'+\left[ \frac{z^2\sigma^2}{(z-b)^2} + \frac{\mu_0^2z}{(z-b)} + \frac{\lambda}{z(z-b)} \right] Z=0.
        \label{radialmelhor}
    \end{equation}
    By performing a $S$-homotopic transformation \cite{ronveaux}
    \begin{equation}
        Z(z)=z^A(z-b)^B \exp\left(Cz\right) f(z),
    \end{equation}
one can find that by setting the exponents $A=0$, $B=ib\sigma$, $C=i\sqrt{\mu_0^2+\sigma^2}$ and with a new variable $x'=1-z/b$, equation \eqref{radialmelhor} transforms into a confluent Heun equation (CHE), which has the canonical form \cite{ronveaux} 
    \begin{equation}
        \frac{d^2\Tilde{F}}{dx'^2}+\left( \frac{\gamma}{x'} + \frac{\delta}{x'-1} + \varepsilon  \right)\frac{d\Tilde{F}}{dx'}+\left[\frac{\alpha x'-q}{x'(x'-1)}\right]\Tilde{F}=0
    \end{equation}
    whose solutions are given by the confluent Heun functions (CHF)
    \begin{equation}
        \Tilde{F}(x)=c_1Hc(q,\alpha,\gamma,\delta;x') + c_2x^{1-\gamma}Hc(q',\alpha',\gamma',\delta';x').
    \end{equation}
where the parameters of the CHF are
    \begin{equation}
    \begin{aligned}
        q&=\sqrt{\mu_0^2+\sigma^2}(-bi+2b^2\sigma)-(\mu_0^2+ 2\sigma^2)b^2-ib\sigma-\lambda, \\
        \alpha&=\sqrt{\mu_0^2+\sigma^2}(-2bi+2b^2\sigma)-(\mu_0^2+ 2\sigma^2)b^2, \\
        \gamma&=1+2ib\sigma, \\
        \delta&=1, \\
        \varepsilon&=-2ib\sqrt{\mu_0^2+\sigma^2}.
    \end{aligned}
    \end{equation}
    and for the second solution
    \begin{equation}
    \begin{aligned}
        q'&=q+(1-\gamma)(\varepsilon-\delta),\\
        \alpha'&=\alpha+(1-\gamma)\varepsilon,\\
        \gamma'&=2-\gamma,\\
        \delta'&=\delta,\\
        \varepsilon'&=\varepsilon.
    \end{aligned}
    \end{equation}
    Therefore, the radial-like solution in the $z$ variable can be written as
    \begin{equation}
    \begin{split}
        Z(z)&=(z-b)^{ib\sigma}\exp{\left(iz\sqrt{\mu_0^2+\sigma^2}\right)}\bigg\{ c_1 Hc\left(q,\alpha,\gamma,\delta;1- \frac{z}{b}\right) \\
        &+ c_2\left( 1-\frac{z}{b} \right)^{1-\gamma}Hc\left(q',\alpha',\gamma',\delta';1-\frac{z}{b}\right) \bigg\}.
    \end{split}
    \label{solucaoradial}
    \end{equation}
    In order to compute the energy eigenvalues $\sigma$, we can make use of the process explained by Schmidt and Pereira \cite{schmidt}, that applied the methodology presented by Kristensson \cite{kristensson}. The well-known Leaver's method \cite{metodo-leaver} relies on the study of continued fractions, which on their turn are quite delicate numerically. Our method is simpler and faster. Solving the transcendental equation given by (equation (A5) of \cite{schmidt}) the condition of vanishing determinant of the tridiagonal matrix (which is obtained by the three-term recurrence relation coefficients that the CHE provides) one can find these eigenvalues. In practice, we use the software Mathematica to solve the transcendental equation with a suitable precision, and the first results for $b= 1/2$, $\mu_0= 2$ and $\lambda= -2$ can be read from the table \ref{tabelasigma}.

    \begin{table}[!htb]
    \centering
    \begin{tabular}{|c|c|}
    \hline
    $\sigma_n$  & Value                                  \\ \hline
    $\sigma_1$  & $-4.253i$                        \\ \hline
    $\sigma_2$  & $-1.146 - 4.067i$ \\ \hline
    $\sigma_3$  & $-3.404 - 2.489i$                     \\ \hline
    $\sigma_4$  & $-2.283 - 3.493i$                       \\ \hline
    $\sigma_5$  & $1.454 + 3.650i$   \\ \hline
    $\sigma_6$  & $0.359 + 3.495i$                   \\ \hline
    $\sigma_7$  & $2.667i$                    \\ \hline
    $\sigma_8$  & $1.544i$                  \\ \hline
    $\sigma_9$  & $0.279i$                  \\ \hline
    $\sigma_{10}$ & $17.29 -0.616i$       \\ \hline
    \end{tabular}
    \caption{First 10 energy eigenvalues obtained with Mathematica. These complex energy eigenvalues may indicate the presence of quasi-normal modes in this system. We used Planck units and $b=1/2, \mu_0=2,$ and $\lambda=-2$. }
    \label{tabelasigma}
    \end{table}

\section{Hawking spectrum and quasi-normal modes}\label{sec:Hawking}

    According to Sannan \cite{sannan}, the presence of an event horizon is the essence of thermal radiation experienced by observers. Since our radial solution has one located at $z=b$ we can apply the Damour-Ruffini-Sannan method \cite{sannan,damour,vieira,dalpra} to extract information about the radiation spectrum emitted by this type of source and the Hawking temperature of the system. 
    
    The solution \eqref{solucaoradial} near the event horizon behaves as
    \begin{equation}\label{prox_horiz}
        Z(z)\approx c_1(z-b)^{ib\sigma} + c_2(z-b)^{-ib\sigma} = Z_1(z) + Z_2(z)
    \end{equation}
    where all the constant terms are absorbed by the constants $c_1$ and $c_2$. This solution can be interpreted as waves going towards the event horizon ($Z_2$) and outgoing waves from the event horizon to infinity ($Z_1$), both in the region $z>b$. The latter is the case of interest to this analysis. 
    
    To introduce null coordinates we may first define the tortoise-like coordinate in this spacetime
    \begin{equation}
        z^*=\int \left( \frac{b}{z}-1 \right)^{-1}dz=-z -b\ln|z-b|.
    \end{equation}
    Therefore, the null coordinate $u$ is
    \begin{equation}
        u=t-z^*
    \end{equation}
    \begin{equation}
        t= u -z -b\ln|z-b|.
    \end{equation}
    Bringing back the temporal portion of the wavefunction and writing it with the new coordinate $u$ we have
    \begin{equation}
        \mathcal{Z}_1(u,z>b)=c_1\exp{[-i\sigma(u+z^*)](z-b)^{ib\sigma}}
    \end{equation}
    \begin{equation}
        \mathcal{Z}_1(u,z>b)=c_1\exp{[-i\sigma(u-z)](z-b)^{2ib\sigma}}.
    \end{equation}
    
    To obtain the decay rate is necessary to analytically continue the function $\mathcal{Z}_1(u,z>b)$ to the region $z<b$. This is possible by means of a rotation of $-\pi$ in the lower half $z$ complex plane
    \begin{equation}
        \mathcal{Z}_1(u,z<b)=c_1\exp{[-i\sigma(u-z)]\cdot[(b-z)e^{-i\pi}]^{2ib\sigma}}
    \end{equation}
    \begin{equation}
        \mathcal{Z}_1(u,z<b)=c_1\exp{[-i\sigma(u-z)](b-z)^{2ib\sigma}}e^{2\pi b\sigma}.
    \end{equation}
    Hence, the decay rate is given by
    \begin{equation}\label{razaoHawking}
        \Gamma= \left| \frac{\mathcal{Z}_1(u,z>b)}{\mathcal{Z}_1(u,z<b)} \right|^2=e^{-4\pi b \sigma}
    \end{equation}
    where $\sigma$ is the energy of the particle emitted. With this result at hands, it is possible to calculate the Hawking radiation spectrum
    \begin{equation}
        N=\frac{\Gamma}{1-\Gamma}=\left( e^{4\pi b \sigma}-1 \right)^{-1}
        \label{radiacaohawking}
    \end{equation}
    and by a simple comparison between \eqref{radiacaohawking} and the blackbody radiation spectrum
    \begin{equation}\
        N=\frac{1}{\exp\left(\frac{E}{k_BT}\right)-1}, \label{hawking-spectrum}
    \end{equation}
    one identifies the so-called Hawking temperature of the tachyon singularity system
    \begin{equation}
        T_H=\frac{1}{4\pi k_B b}.
    \end{equation}

\begin{figure}[!htb]
        \centering
        \includegraphics[width=.7\linewidth]{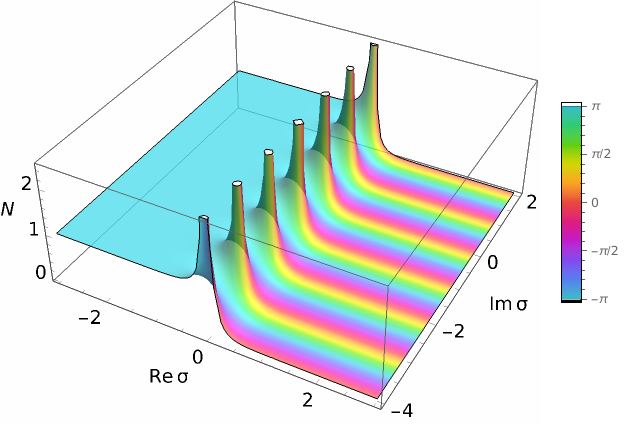}
        \caption{Hawking radiation spectrum $N$ as a function of the complex energy $\sigma$, given by \eqref{radiacaohawking}, for the mass parameter $b=1/2$ and considering that the energy $\sigma \in \mathbb{C}$. We observe that there are poles whenever $\sigma = i n/(2b)$, where $n$ is an integer.}
        \label{plotHawking}
    \end{figure}

    With the aim of obtaining better insights and visualization of the phenomena discussed, we may express visually the radiation spectrum given by equation \eqref{radiacaohawking}. We should bear in mind the fact that the energy may be a complex number, since we have already calculated it in table \ref{tabelasigma} and argued that this indicates the presence of quasi-normal modes of the tachyonic singularity. We plot the complex radiation spectrum in figure \ref{plotHawking}, in which we are able to see the exponential nature of the real part of the spectrum and the oscillatory behavior of its imaginary part, which diverges every time the factors in the denominator cancel each other.
    
    Another key aspect of the physics of black holes is the presence of quasi-normal modes. These modes are defined as a complex set of frequencies related to solutions of the perturbation equations of the singularity in matter that obey appropriate boundary conditions \cite{chandrasekhar}. The idea of quasi-normal modes can be elucidated if one imagines the black hole (or a gravitational singularity) as an immense bell that, long after being rung, sounds its last notes \cite{chandrasekhar_livro} that propagate through the spacetime fabric. In our case, the scalar particle is the perturbation to the metric, and the boundary conditions are imposed on the radial solution so that it is finite at the horizon $z=b$ and the function is well behaved at infinity \cite{vieiraquasinormal}. 
    
    The interpretation of equation \eqref{prox_horiz} shows that the first condition is already satisfied. For the second condition, it is required that \eqref{solucaoradial} has a polynomial form. To achieve this, two conditions are necessary, as carefully explained in the appendix of \cite{schmidt}. The confluent Heun's functions parameters must obey, 
    \begin{equation}\label{condicaoalpha}
        \alpha=-\varepsilon(m+s-1)
    \end{equation}
    where for the first solution $s_1=0$ and for the second $s_2=1-\gamma$, and the vanishing determinant of the tridiagonal matrix
    \begin{equation}\label{determinante}
        D_m=0.
    \end{equation}
    This generates {\it two consecutive} null coefficients in the three-term recurrence relation that the confluent Heun equation satisfies \cite{kristensson}, resulting in a polynomial solution with $m$ terms. Finally, the complex frequencies are extracted from equation \eqref{determinante} combined with \eqref{condicaoalpha} and presented in table \ref{tabela_quasinormal}. We have $n$ modes for each different value of $m$. We use the same parameters $b= 1/2$, $\mu_0= 2$ and $\lambda= -2$ to make it clear that the values are different from the energy eigenvalues in table \ref{tabelasigma}.

    \begin{table}[]
\begin{tabular}{|c|c|c|c|c|}
\hline
             & $m=1$            & $m=2$            & $m=3$            & $m=4$            \\ \hline
$\sigma_{1}$ & $0.509-0.152i$ & $3.388i$       & $-3.164i$      & $19.527i$      \\ \hline
$\sigma_{2}$ &    ---            & $-2.297i$      & $2.537i$       & $-4.111i$      \\ \hline
$\sigma_{3}$ &    ---        & $0.376-0.760i$ & $-2.357i$      & $-3.347i$      \\ \hline
$\sigma_{4}$ &    ---            & $0.207-0.180i$ & $0.435-1.101i$ & $-2.403i$      \\ \hline
$\sigma_{5}$ &    ---            &     ---           & $0.182-0.788i$ & $2.302i$       \\ \hline
$\sigma_{6}$ &    ---            &     ---           & $0.183-0.224i$ & $0.516-1.439i$ \\ \hline
$\sigma_{7}$ &    ---            &     ---           &     ---           & $0.227-1.277i$ \\ \hline
$\sigma_{8}$ &    ---            &     ---           &     ---           & $0.152-0.832i$ \\ \hline
$\sigma_{9}$ &    ---            &     ---           &     ---           & $0.159-0.254i$ \\ \hline
\end{tabular}
\caption{Quasi-normal modes of the tachyonic singularity. Here we use Planck units where $\hbar=c=G=1$, and we consider $b=1/2, \mu_0=2, $ and $\lambda=-2$. }
\label{tabela_quasinormal}
\end{table}

\section{Analogue model for radial dynamics}\label{sec:Analogo Radial}

The ionic hydrogen molecule, $H_2^+$, is a system composed of two protons and one orbiting electron. Because the protons are much heavier than the electron, one can safely model this system as if the two protons are stationary and the single electron moves under the influence of the two Coulomb interactions due to the protons. A suitable coordinate system \cite{wilson,Lay_2024,baber-hasse,leaver} to describe this is the prolate spheroidal coordinate system, defined as%
\begin{equation} \label{prolate}
    \begin{aligned}
        X &= d\sinh\eta \sin\theta \cos\phi, \\
        Y &= d\sinh\eta \sin\theta \sin\phi,\\
        Z &= d\cosh\eta\cos\theta,
    \end{aligned}
\end{equation}
where $2d$ is the distance between the two protons, and constant $\eta$ defines a spheroidal shell. The Schr\"{o}dinger equation reads
\begin{equation}
    \nabla^2\psi + (E - 2V)\psi = 0,
\end{equation}
where $V$ is the electric potential, which can be defined as
\begin{equation}\label{potential}
    V(r_1,r_2) = \frac{N_1 N_2}{R} - \frac{N_1}{r_1} - \frac{N_2}{r_2}.
\end{equation}

\begin{figure}[!htb]
        \centering
        \includegraphics[width=.6\linewidth]{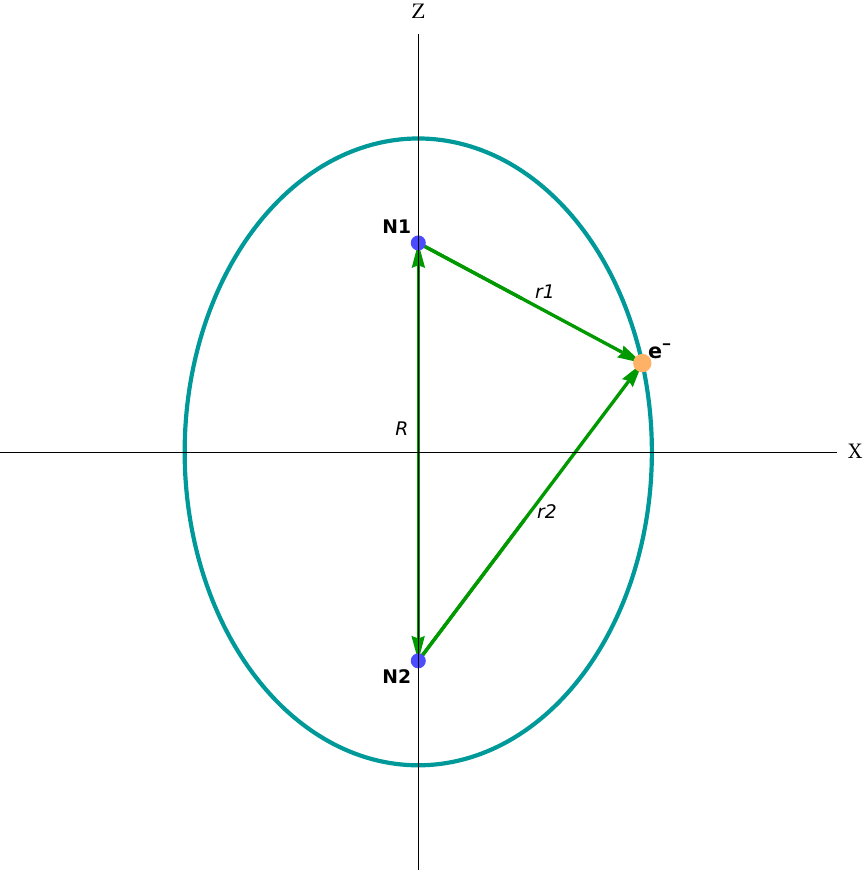}
        \caption{Electron interacting with two static nuclei, with charges $N_1$ and $N_2$. The potential is given by equation (\ref{potential}) and the suitable coordinate system is the prolate spheroidal (\ref{prolate}).}
        \label{plotH2+}
    \end{figure}

We have defined $r_1$ and $r_2$ as the distance between the electron to the nuclei $1$ and $2$, respectively, while $R=2d$ is the distance between the two nuclei, and $N_1$, $N_2$ their respective charges, as multiples of the elementary charge, see figure \ref{plotH2+}. The units are such that $\hbar = m^* = 4\pi\epsilon_0 = q_e = 1$, where $m^*$ is the electron's mass, $q_e$ is its charge and $\epsilon_0$ is the vacuum permittivity. Because it is the hydrogen molecule, $N_1 = N_2 = 1$, and the prolate spheroidal coordinates naturally arise as the most fitting coordinate system to work with. Decoupling Schr\"{o}dinger equation using simple separation of variables as $\psi(\eta,\theta,\phi) = e^{i n \phi}\Theta(\theta)y(\eta)$ and using $z = \cosh\eta$, we get
  \begin{multline}
         y''+\left( \frac{1}{z+1} +\frac{1}{z-1} \right)y' +\Bigg[ - \frac{n^2}{4(z-1)^2} - \frac{n^2}{4(z+1)^2} + \frac{-2\lambda'^2+2\kappa-2\mu+n^2}{4(z-1)} \\ 
         + \frac{2\lambda'^2+2\kappa+2\mu-n^2}{4(z+1)} 
        -\lambda'^2 - \frac{R^2}{2}\frac{V_\text{ext}}{z^2 - 1}\Bigg]y=0
        \label{h2+}
   \end{multline}
   where $\mu$ is a separation constant, $n$ is an integer azimuthal quantum number. The constants $\lambda'$ and $\kappa$ are defined as 
    \begin{equation}
        \lambda'^2= - \frac{2m^* d^2}{\hbar^2}E',
    \end{equation}
and
    \begin{equation}
         \kappa=\frac{4m^*  q_e^2d}{\hbar^2}=\frac{2d}{a_0}.
    \end{equation}
Here, $E'$ is the energy of the lonely electron, defined as $E' = E - N_1 N_2/R$, $a_0$ is the Bohr radius and $V_\text{ext}$ is an external potential.

With a new variable $x=(1+z)/2$, and defining $y(z)=y(2x-1)=f(x)$, equation \eqref{h2+} can be written in a more convenient form,
    \begin{multline}
        f''+\left( \frac{1}{x} +\frac{1}{x-1} \right)f' +\Bigg[ - \frac{n^2}{4x^2} - \frac{n^2}{4(x-1)^2} + \frac{\lambda'^2+\kappa+\mu-n^2/2}{x} \\+ \frac{-\lambda'^2+\kappa-\mu+n^2/2}{x-1} -4\lambda'^2 - \frac{R^2}{2}\frac{V_\text{ext}}{x(x-1)}\Bigg]f=0.
                \label{h2+melhor}
    \end{multline}
    So as to be able to reproduce the dynamics arising from the scalar particle in the tachyonic background in the laboratory $H_2^+$ system, we must introduce a non-central external potential $V_{ext}(x)$ such that
    \begin{equation}\label{potencial-externo-radial}
        V_\text{ext}(x)= \left(\frac{-R^2}{2x(x-1)}\right)^{-1}\left(\frac{-2\kappa }{(x-1)}+\frac{n^2}{4x^2}+\frac{4V_0 x^2+n^2}{4(x-1)^2}+\frac{V_1x}{(x-1)} + 4\lambda'^2\right).
    \end{equation}

    This brings equation \eqref{h2+melhor} to the form 
    \begin{equation}
        f''+\left( \frac{1}{x} +\frac{1}{x-1} \right)f'+\left[ \frac{V_0x^2}{(x-1)^2} +\frac{V_1x}{(x-1)} + \frac{-\lambda'^2+\kappa-\mu+n^2/2}{x(x-1)}\right]f=0.
        \label{h2+potencial}
    \end{equation}

Now, let us reconsider the radial equation from the tachyon problem. With a change that rescales our initial variable from equation \eqref{radialmelhor} $z=bz'$, and $Z(z)=g(z')$, we have
    \begin{equation}
        g''+\left( \frac{1}{z'}+\frac{1}{z'-1} \right)g'+\left[ \frac{b^2z'^2\sigma^2}{(z'-1)^2} + \frac{\mu_0^2b^2 z'}{(z'-1)} + \frac{\lambda}{z'(z'-1)} \right] g=0.
        \label{radialescala}
    \end{equation}
Therefore, the correspondence between \eqref{h2+potencial} and  \eqref{radialescala} is manifest if one is able to perform a fine tuning in the parameters of the laboratory system as $V_0=b^2\sigma^2,$ $V_1=\mu_0^2b^2,$ and $-\lambda'^2+\kappa-\mu+n^2/2=\lambda_{t}$. However, in the laboratory system, Wilson \cite{wilson} used atomic units, and we are using Planck units. To make a direct comparison, we rewrite these results in SI units, obtaining
    \begin{equation}\label{analogia-parametros-eq-radial}
    \begin{split}
        V_0=\frac{G^2b^2\sigma^2}{2m^* c^6 d^2},\qquad         V_1=\frac{\mu_0^2 G^2 b^2}{4m^* d^2},\qquad
        -\lambda'^2+\kappa-\mu+ \frac{n^2}{2}=\lambda_{t},
    \end{split}
    \end{equation}
where $2d$ is the distance between nuclei and we call $\lambda_t$ the separation constant from the tachyon system to avoid ambiguity. Observe that in the above equations the LHS refers to physical quantities of the laboratory system, while the RHS has physical quantities of the target system. So, there is an explicit one-to-one mapping between the physical quantities of the target and laboratory systems.

In fact, if we reintroduce the constants $\hbar, G,$ and $c$ it is possible to estimate that for a tachyon with a mass of the order of $10^5 kg$, and the complex energy of the test particle being (pure imaginary) of the order of $46i J$, the coupling constant $V_0$ of the potential must be of the order of $0.1\% - 1\%$ of the coupling constant of the Coulomb potential. And, from this conclusion, considering $\mu_0$ to be of the order of $10^{-27}kg$ the constant $V_1$ becomes negligible. Thus, for practical purposes, the laboratory system that replicates the radial dynamics of a test particle interacting with the gravitational field of a tachyon can be modeled as an electron interacting with an $H_2^+$ molecule subject to an external potential \eqref{potencial-externo-radial} where $V_1\approx 0$ and $V_0 \sim 10^{-11} J\cdot m $.

Furthermore, with the analogy between the equations \eqref{h2+potencial} and \eqref{radialescala} established, it is also possible to recognize that the singularities of our differential equation \eqref{radialescala} store valuable information about the system in question. In other words, the singular points of the equation are the physical points of interest of the problem, namely, the event horizon and the curvature singularity. With that remark, our interpretation is that the point that plays the role of the event horizon in the analogous system is given by the singularity $x=1$ in equation \eqref{h2+potencial}, where we find $\eta=0$ in the original coordinate system. In the prolate spheroidal coordinates, $\eta=0$ represents the line segment connecting the two fixed protons of the molecule. On the other hand, the curvature singularity is represented by $x=0$ in equation \eqref{radialescala}, and this point asks for a value of $\eta$ that is outside the physical domain, since $\cosh\eta =0$ is not satisfied by any real $\eta$.

As a consequence of this perspective, the entire analysis and interpretation of section \ref{sec:Hawking} regarding the behavior of the wavefunction in the neighborhood of the event horizon can now be applied to the electron's wavefunction. Hence, it is expected the observation of a net flux of outgoing probability density of the wave function in the vicinity of the segment $\eta=0$, according to the equation \eqref{razaoHawking}. This is due to the fact that the applied external potential is non-central, occasioning a break in the expected symmetry of the problem with respect to the level surfaces $\eta=const$. This flux depends on the square root of the intensity of the coupling constant $V_0$ and is analogous to Hawking radiation in the system.

One more time, following the analysis in section \ref{sec:Hawking}, we may now consider the analogue quasi-normal modes. With the one-to-one relations \eqref{analogia-parametros-eq-radial} and the conditions \eqref{condicaoalpha} and \eqref{determinante} imposed on the wavefunction, it is expected that, at an appreciable distance from the external and Coulomb potentials, the observed energy levels for the electron will replicate the quasi-normal modes presented in table \ref{tabela_quasinormal} of the scalar particle in the tachyon background. 

\section{Analogue model for angular dynamics}\label{sec:Analogo Angular}

Consider the radial equation from the hydrogen molecule ion in the original coordinate system
    \begin{equation}
        f''+ f'\coth{\eta} + \left( -\lambda'^2\cosh^2{\eta}+\kappa\cosh{\eta}-\mu -\frac{n^2}{\sinh^2{\eta}} + \frac{R^2}{2}V_\text{ext}\cosh^2{\eta} \right)f=0.
    \end{equation}
The introduction of a carefully tailored external potential,
\begin{equation}
        V_\text{ext}(\eta)=\frac{2}{R^2}\left(\lambda'^2-\frac{\kappa}{\cosh{\eta}}\right),
        \label{vext_analogo_angular}
    \end{equation}
    makes the $H_2^+$ radial equation take the form, 
    \begin{equation}
        f''+ f'\coth{\eta} + \left( -\mu -\frac{n^2}{\sinh^2{\eta}} \right)f=0.
        \label{angularanalogo}
    \end{equation}
    Back to the tachyon system, we remember the angular equation \eqref{angular}, that can be rewritten as 
    \begin{equation}
        R''+R' \coth{r} + \left( \lambda -\frac{m^2}{\sinh^2{r}} \right)R=0.
        \label{angularmelhor}
    \end{equation}

Thus, there is a one-to-one correspondence established between equations \eqref{angularanalogo} and \eqref{angularmelhor}. It is straightforward to carry out a simple transcription --- a dictionary --- in the laboratory parameters,

    \begin{equation}\label{analogia-parametros-eq-angular}
    \begin{split}
        n=m_t, \qquad         \mu=-\lambda_t,
    \end{split}
    \end{equation}
    where, again, the subscript $t$ indicates the quantity that belongs to the particle in the tachyon background system.

    We must emphasize that the simplicity of equation \eqref{analogia-parametros-eq-angular} is somewhat remarkable, both for the beauty of the theory and for experimental purposes. If one is able to manufacture such external potential \eqref{vext_analogo_angular}, then there is a one-to-one relation between the numerical value of the angular quantum numbers of the electron and the test particle in the tachyon background. 

\section{Conclusions}\label{sec:Conclusao}

    We presented the exact solution for the Klein-Gordon equation in the background of the AII metric, the tachyon spacetime metric. The angular-like and the radial-like solutions are given by generalized Legendre and confluent Heun functions, from which we are able to extract values for the separation constant and energy eigenvalues. We also applied the Damour-Ruffini-Sannan method to obtain the Hawking radiation spectrum and temperature associated with the metric's event horizon. In addition, we calculated the quasi-normal modes of the tachyonic singularity employing a technique that consists of imposing boundary conditions in the radial part of the Klein-Gordon equation solution. From the perspective of the analogue gravity research programme, we proposed a quantum model that imitates the gravitational one, the $H_2^+$ molecule. 
    
    Regarding the physical realization of such model, it is important to note that the potential \eqref{potencial-externo-radial} is energy-dependent. This can be implemented \cite{sasada2008calculation,rotter2009non} in an open quantum system, which offers the required boundary conditions necessary for the spectroscopy of quasinormal modes. Then, the decay of excited states of $H_2^+$ is proportional to the imaginary part of the frequency, while the oscillation frequency of the molecule can be measured via spectroscopy and is proportional to the energy of the scalar particle in the tachyonic background. Recently, the technique for the control of a single $H_2^+$ ion has been achieved \cite{single-h2-ion}. This result leads to novel development in the high-precision spectroscopy of the molecule, and it could be a way to achieve the implementation of our analogue model. However, we should emphasize that the physical assumptions regarding the external potential and constant adjustments may turn to be a substantial experimental bottleneck for current techniques and laboratory procedures. Nevertheless, the concepts presented in this work have the potential to generate useful ideas for a diverse audience of readers interested in interdisciplinary topics, such as the overlap between molecular physics and general relativity.
    
    In summary, we were able to match the exact same equations of motion in the quantum system and in the tachyonic system, constructing a map between the physical parameters of both systems. For the radial problem this mapping is given by \eqref{analogia-parametros-eq-radial}, and for the angular problem the mapping is given by \eqref{analogia-parametros-eq-angular}.

\begin{acknowledgments}
The authors gratefully acknowledge CNPq (grant numbers 140471/2022-7 and 309052/2023-8) and CAPES (grant number 88887.990311/2024-00) for financial support. 

The manuscript has no associated data.

\end{acknowledgments}
    
\bibliography{ref}

\end{document}